\documentclass[sigconf, nonacm]{acmart}

\AtBeginDocument{%
  \providecommand\BibTeX{{%
    \normalfont B\kern-0.5em{\scshape i\kern-0.25em b}\kern-0.8em\TeX}}}

\copyrightyear{2024}
\acmYear{2024}
\setcopyright{CC}
\acmConference[ASE '24]{39th IEEE/ACM International Conference on Automated Software Engineering }{October 27-November 1, 2024}{Sacramento, CA, USA}
\acmBooktitle{39th IEEE/ACM International Conference on Automated Software Engineering (ASE '24), October 27-November 1, 2024, Sacramento, CA, USA}
\acmDOI{10.1145/3691620.3695515}
\acmISBN{979-8-4007-1248-7/24/10}

\usepackage{algpseudocode}
\usepackage{pgfplots}
\usepackage{multirow}
\usepackage[normalem]{ulem}
\usepackage{xcolor,colortbl}
\usepackage{hhline}
\usepackage{subcaption}
\usepackage[most]{tcolorbox}
\usepackage{xspace}
\usepackage{xcolor}

\pgfplotsset{compat=1.3}
\usetikzlibrary{patterns}

\definecolor{grey}{gray}{0.9}
\definecolor{green-draw}{HTML}{97D077}
\definecolor{blue-draw}{HTML}{7EA6E0}

\newcolumntype{P}[1]{>{\columncolor{grey}}p{#1}}
\newcolumntype{C}[0]{>{\columncolor{grey}}c}

\def\papername{LeanBin\xspace}

\usepackage{hyphenat}

\newtcolorbox[auto counter]{summary}[1][]{
  title={\bfseries Summary RQ\thetcbcounter},
  enhanced,
  drop shadow={black!50!white},
  coltitle=black,
  top=0.15in,
  attach boxed title to top left={xshift=1.5em,yshift=-\tcboxedtitleheight/2},
  boxed title style={size=small,colback=pink},
  #1
}

\begin{document}

\title{\papername: Harnessing Lifting and Recompilation to Debloat Binaries}

\author{Igor Wodiany}
\email{igor.wodiany@manchester.ac.uk}
\orcid{0000-0002-7682-5581}
\affiliation{%
  \department{Department of Computer Science}
  \institution{University of Manchester}
  \streetaddress{Oxford Road}
  \city{Manchester}
  \country{United Kingdom}
  \postcode{M13 9PL}
}

\author{Antoniu Pop}
\email{antoniu.pop@manchester.ac.uk}
\orcid{0000-0002-7715-4281}
\affiliation{%
  \department{Department of Computer Science}
  \institution{University of Manchester}
  \streetaddress{Oxford Road}
  \city{Manchester}
  \country{United Kingdom}
  \postcode{M13 9PL}
}

\author{Mikel Luj\'an}
\email{mikel.lujan@manchester.ac.uk}
\orcid{0000-0002-0842-1083}
\affiliation{%
  \department{Department of Computer Science}
  \institution{University of Manchester}
  \streetaddress{Oxford Road}
  \city{Manchester}
  \country{United Kingdom}
  \postcode{M13 9PL}
}


\begin{abstract}

To reduce the source of potential exploits, \emph{binary debloating} or \emph{specialization} tools are used to remove unnecessary code from binaries. This paper presents a new binary debloating and specialization tool, \papername, that harnesses lifting and recompilation, based on \emph{observed execution traces}. The dynamically recorded execution traces capture the required subset of instructions and control flow of the application binary for a given set of inputs. This initial control flow is subsequently augmented using \emph{heuristic-free} static analysis to avoid excessively restricting the input space. The further structuring of the control flow and translation of binary instructions into a subset of C enables a lightweight generation of the code that can be recompiled, obtaining LLVM IR and a new debloated binary. Unlike most debloating approaches, \papername enables both binary debloating of the application and shared libraries, while reusing the existing compiler infrastructure. Additionally, unlike existing binary lifters, it does not rely on potentially unsound heuristics used by static lifters, nor suffers from long execution times, a limitation of existing dynamic lifters. Instead, \papername combines both heuristic-free static and dynamic analysis. The run time of lifting and debloating SPEC CPU2006 INT benchmarks has a geomean of 1.78$\times$, normalized to the native execution, and the debloated binary runs with a geomean overhead of 1.21$\times$. The percentage of gadgets, compared to the original binary, has a geomean between 24.10\% and 30.22\%, depending on the debloating strategy; and the code size can be as low as 53.59\%. For the SQLite use-case, \papername debloats a binary including its shared library and generates a debloated binary that runs up to 1.24$\times$ faster with 3.65\% gadgets.

\end{abstract}

\ccsdesc[500]{Software and its engineering~Software maintenance tools}
\ccsdesc[500]{Software and its engineering~Source code generation}
\ccsdesc[500]{Security and privacy~Software reverse engineering}
\ccsdesc[300]{Security and privacy~Software security engineering}

\keywords{Binary debloating, Binary specialization, Binary lifting, Control-flow recovery, Recompilation, Heuristc-free static analysis}


\maketitle

\textbf{NOTE}

\noindent
This is an extended version of the paper accepted to 39th IEEE/ACM International Conference on Automated Software Engineering (ASE ’24), October 27 - November 1, 2024, Sacramento, CA. The definitive Version of Record will be published in ACM Digital Library: \url{https://doi.org/10.1145/3691620.3695515}.

\section{Introduction}

To reduce potential exploits, \emph{binary debloating} or \emph{specialization} tools, are used to remove unnecessary code from binaries. For example, an average program uses only around 5\% of the whole \emph{libc} \cite{piece-wise}. Debloating can be done either on the application source code or directly at the binary level, and often needs to be applied to production-optimized stripped binaries (no source code, no symbol table). When working only with binaries, the debloating tools can \textit{lift} the executable binary, or parts of, into a well-supported programming language, such as C, or into an intermediate representation (IR). This IR can either be low-level and designed specifically for binary analysis \cite{binary-ir} or high-level and general-purpose, e.g., LLVM IR \cite{llvm}. In this way, binary lifting can complement, and facilitate the reuse of analyses and transformations in existing tools.

However, correctly recovering a high-level representation from existing binaries that are optimized, stripped or obfuscated remains a challenge. The majority of binary lifters rely on static heuristics to recover the control flow of the application. Yet, the precision of those heuristics is limited, and static lifters often impose a number of restrictions on binaries \cite{secondwrite-1}. Dynamic lifting overcomes the limitations associated with static lifters, offering a compelling alternative for precise and correct lifting that does not rely on heuristics. However, fully dynamic lifting has an intrinsic execution time overhead, as the lifting happens as an extra computation during the binary execution. For example, a recent dynamic lifter, BinRec \cite{binrec}, reports that lifting of a single SPEC 2006 benchmark takes on average 50 hours (about 2 days) as it uses S\textsuperscript{2}E \cite{s2e} that emulates the entire environment. Moreover, the purely dynamic approach may not lift all relevant parts of the application, excessively restricting the application input space.

This paper presents a new lifter-based debloating approach, and the tool \papername, that combines dynamic control-flow tracing based on observed execution with heuristic-free static lifting. Our approach relies on a lightweight tracer to record control flow, executed basic blocks and their boundaries, with low overhead. This run-time information, an \emph{observed execution trace}, captures the required subset of instructions of the binary for a given set of inputs. Observed execution traces represent expected execution paths, and are further augmented with heuristic-free static analysis, to avoid excessively restricting the application input space. Subsequently, the recovered CFG is structured and statically lifted into a subset of C. The lifted C can be then used to generate LLVM IR, or compiled into a debloated binary which may also include the executed parts of shared libraries.

One of the key advantages of our approach is to combine the precision offered by fully dynamic solutions with faster lifting times that are more common with static lifters, by using both heuristic-free static and dynamic lifting techniques. By doing this, \papername fills the gap in the design space between static and dynamic lifters (see Section \ref{sec:rw}). \papername makes a limited set of assumptions, enabling lifting of highly optimized, position-independent, stripped, dynamically linked binaries. The robustness of \papername is evaluated with the SPEC CPU2006 INT benchmark suite, with a use-case based on SQLite (Section~\ref{sec:evaluation}) and with multi-threaded Phoenix benchmarks~\cite{phoenix}. Although the implementation targets AArch64, the approach described is architecture-independent. The only assumption is that we are able to identify which binary instructions are branches, as in principle, the control flow of applications in all major architectures is compiled into direct and indirect branches.

To summarize, we make the following contributions:
\begin{itemize}
    \item \papername is the first debloater, implemented for multi-threaded AArch64 binaries, that uses hybrid lifting, a novel approach that combines heuristic-free static and dynamic analysis.
    \item \papername can correctly debloat optimized binaries from the SPEC CPU2006 INT benchmark suite. The lifting overhead is as small as 1.05$\times$ with a 1.78$\times$ geomean, normalized to native execution.
    \item We show that the number of gadgets, compared to the original binary, has a geomean between 24.10\% and 30.22\%, depending on the debloating strategy, going down to 16.65\% when using ARM PAC pointers. The size reduction has a geomean of 53.59\% for a single execution trace.
    \item For an SQLite-based use-case, by debloating a binary together with its dependency, \papername generates a debloated binary that runs up to $1.24\times$ faster with as little as 4.47\% gadgets.
    \item The implementation of the tool is open source and available on GitHub: \url{https://github.com/IgWod/leanbin.git}.
\end{itemize}

\section{Design and Implementation}
\label{sec:design}

\begin{figure*}[t]
    \centering
    \includegraphics[scale=0.48]{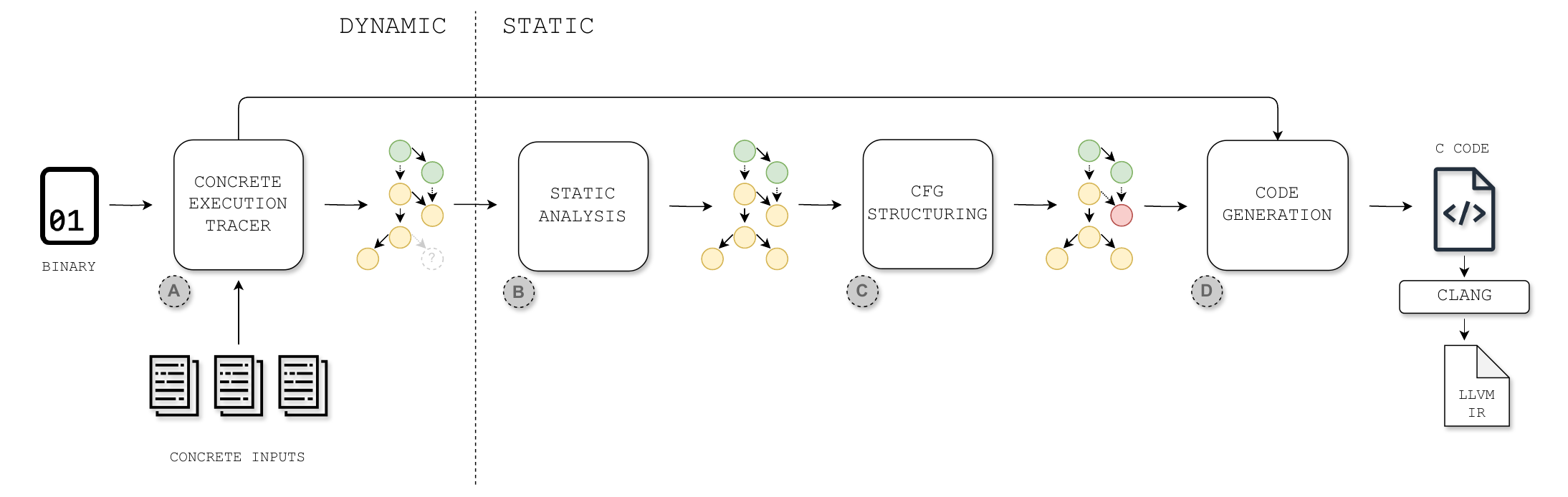}
    \caption{Debloating steps of \papername: (A) Dynamic concrete execution tracing; (B) Static heuristic-free analysis of the control flow; (C) Control-flow structuring; (D) Instructions lifting and code generation.}
    \Description{The figure illustrates the flow of data and information within the system, divided into two sections: dynamic (on the left) and static (on the right). On the dynamic side, the "Concrete Execution Tracer" box has two incoming arrows: one from the binary and another from the inputs. An outgoing arrow from this box leads to a potentially incomplete control-flow graph. This incomplete graph is then passed to the "Static Analysis" box, located on the static side of the figure. The "Static Analysis" box outputs a complete control-flow graph, which is then fed into the "CFG Structuring" box. The "CFG Structuring" box produces a new control-flow graph, which is sent to the "Code Generation" box. The "Code Generation" box also receives the original binary as input. Finally, the "Code Generation" box outputs C code, which is then fed into Clang to produce LLVM IR.}
    \label{fig:system}
\end{figure*}

Figure~\ref{fig:system} presents the overview of \papername. The debloating and lifting steps are split between dynamic and static parts. The dynamic part, (A) concrete execution tracer that collects information about executed basic blocks and indirect branches targets from unmodified binaries. The tracer outputs a control-flow graph (CFG) --- representing the control flow between basic blocks within a function, and among functions --- of the executed application.

The static part consists of: (B) heuristic-free control-flow static analysis to recover unexplored direct branches and expand the CFG; (C) control-flow structuring to refine the recovered graph; (D) instructions lifter that disassembles instructions into Tiny Code Generator (TCG) IR, a QEMU intermediate representation, and then converts them into C statements. Those C statements and the recovered CFG are then combined to generate a \textbf{recompilable} C code. The C code is then passed to \texttt{clang} to generate LLVM IR (or other compilers/IRs), avoiding dependencies on a specific IR version. The following sections discuss in more detail the main components of \papername.

\subsection{Dynamic Tracing}
\label{sec:dynamic-tracer}

To capture dynamic information (referred to as \emph{observed execution traces}) from a running application, the target binary is run under a concrete execution tracer; this is the first step in the debloating process. The purpose of the tracer is twofold: firstly, to record all basic blocks and their boundaries as executed, and secondly to instrument indirect branches for which the target addresses cannot be recovered statically. Indirect branches do not encode the target of the branch within the instruction but instead jump to an address stored in a specified register. This address can be loaded from the memory or be a result of an arithmetic operation, and as such cannot be always determined statically. For direct branches, the target address forms part of the instruction.

To keep the overhead low, we only collect information that cannot be precisely recovered using static analysis. For basic blocks ending with direct branches (\texttt{b}, \texttt{bl}, \texttt{b.cond} on AArch64), we only record the beginning and end of the block, as the target can be calculated directly from the instruction encoding. For indirect branches (\texttt{br}, \texttt{blr}, \texttt{ret}), the tracer inserts a lightweight hash-map-based instrumentation that dynamically records targets of those branches.

\textit{Multi-Threading Support} --- To enable the collection of execution traces efficiently in multi-threaded applications, we record the control flow of each thread in a per-thread private data cache. Once the application terminates, the control-flow information from all the threads is merged. This enables lock-free operation within the tracer internal implementation, while the threads are running --- the multi-threaded behavior of the traced application is unaffected. Moreover, for multi-threaded applications, we use the tracer to dynamically collect entry points to new threads (e.g., calls to \texttt{pthread\_create}).

\textit{Merging Multiple Execution Traces} --- Since the tracer controls the placement of the binary, the address of each basic block (or its offset from where the binary is loaded) can be fixed among runs. By doing that, CFGs generated by multiple runs can be merged. The final CFG consists then of all the unique basic blocks discovered across all runs, and a union of all the edges discovered for every basic block for each run.

\subsection{Heuristic-Free Static Analysis}

As using only observed execution traces collected by the dynamic tracer (Section~\ref{sec:dynamic-tracer})  can excessively restrict the inputs supported by the application, \papername uses heuristic-free static analysis to augment the dynamically discovered CFG. Whereas classical static lifers use analysis that may be unsafe for arbitrary binaries, this work proposes the use of heuristic-free analysis, that only follows unexplored direct control flow, leaving the exploration of indirect branches to the dynamic tracer. The purpose of such an approach is to avoid over-exploring new functionality, which would reduce the benefits of the binary debloating, but allow small variations to inputs, to avoid excessive execution failures, e.g., program failing due to different memory alignment of the same input parameters.

The main idea behind \papername's static analysis is to first follow an unexplored side of a conditional branch and then to continue the exploration of conditional and unconditional control flow until it is no longer possible or desired. This gives rise to two different debloating strategies. In the first strategy, the analysis follows all direct branches (\texttt{b}, \texttt{b.cond} and \texttt{bl} on AArch64) until an indirect branch is reached and the heuristic-free analysis is no longer possible. In the second strategy, the analysis only follows direct branches that are not marked as function calls (\texttt{b} and \texttt{b.cond} only), and that stops when branch-and-link is discovered --- further recovery no longer desired to avoid excessive exploration. In both cases ``dangling'' sequences of branches (sequences that cannot reach the end of the program) are pruned. This happens when the analyzer reaches an indirect branch with no recovered targets. The pruning is done up to the conditional branch that allows reaching the end of the execution.

\begin{figure}[ht]
    \centering
    \includegraphics[scale=0.75]{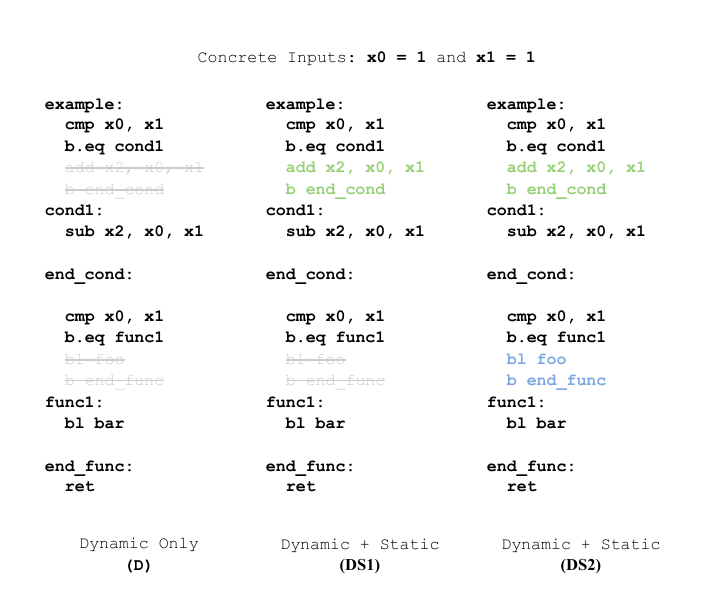}
    \caption{Static CFG analysis --- step (B) in Figure \ref{fig:system}. Instructions in bold represent dynamically discovered code (with \texttt{x0 = 1} and \texttt{x1 = 1}), in \textcolor{green-draw}{green} code discovered with strategy DS1 (static and dynamic analysis), and in \textcolor{blue-draw}{blue} with DS2.}
    \Description{The figure displays three identical assembly code snippets, each labeled with a different title: "Dynamic Only (D)," "Dynamic + Static (DS1)," and "Dynamic + Static (DS2)". At the top of the figure, the text reads: "Concrete Inputs: x0 = 1 and x1 = 1". Each code snippet highlights the specific instructions that are lifted based on the given concrete inputs, using the respective strategies indicated by their titles.}
    \label{fig:static-example}
\end{figure}

The first strategy supports more inputs, at the expense of reduced debloating, whereas the second one is intended to support a wider range of similar inputs without excessively adding new functionality. For the rest of the paper, we refer to those strategies as \textbf{DS2} (includes functions calls) and \textbf{DS1} (excludes function calls) respectively, with \textbf{D} denoting the use of only dynamic traces with static analysis disabled. The example of all described strategies is presented in Figure~\ref{fig:static-example}.

\subsection{Control Flow Structuring}

\begin{figure*}[t]
    \centering
    \includegraphics[scale=0.57]{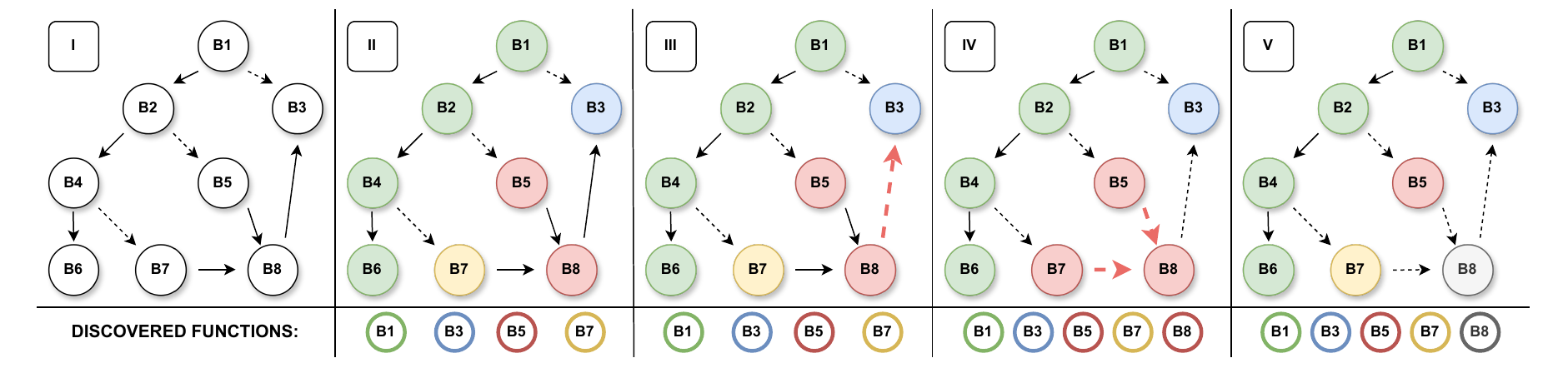}
    \caption{Control flow graph structuring --- step (C) in Figure \ref{fig:system}.}
    \Description{The figure demonstrates the control-flow structuring algorithm using a simple example with eight basic blocks. It illustrates the process of discovering functions and promoting problematic edges.}
    \label{fig:algorithm-example}
\end{figure*}

Lowering the abstraction level, compiler optimizations (e.g., \emph{tail call} optimization) and deliberate obfuscation can change the original structure of the CFG. Such a new structure often does not fit into the structure of high-level representations, such as C and LLVM~IR. For example, no arbitrary control flow transfers between functions are allowed. To enable lifting, the control flow has to be first structured to fit into that high-level representation.

We define the CFG as a directed graph with nodes representing basic blocks, and edges representing the control flow transitions between those blocks. The edge can represent either regular control flow transitions (solid edges in Figure~\ref{fig:algorithm-example}) or function calls (dotted edges). This is to allow the representation of the whole program, with all its functions, as a single graph. The return edges are implicit, as in a standard call graph representation.

We build the initial CFG with information collected by the execution tracer and static analyzer. Firstly, any control-flow edges coming out from basic blocks ending with a branch-and-link instruction (\texttt{bl} and \texttt{blr} on AArch64) are marked as function call edges; step (I) in Figure~\ref{fig:algorithm-example}. An additional edge pointing to the next address, after the function call, is added to indicate that the execution resumes after the call once the function returns. This additional edge is later removed if the called function is shown to never return (e.g., call to \texttt{exit}).

Next, the CFG is split into functions, using initial call edges; step (II). Each basic block (represented as a circle: $B1 \dots B8$) gets assigned an ID (color), and the same ID indicates blocks in the same functions. Such CFG may not be correct, as some edges (e.g., $(B8, B3)$) cross the function boundary.

The problematic control flow edges are then promoted to function calls, effectively performing a function outlining and creating new functions. The promotion happens when the CFG branches to the block that is already called by another block, or when the branch crosses the function boundary. The splitting of CFG into functions and the promotion continues until there are no more edges left to promote.

Red bold arrows indicate edges most recently promoted to calls. Functions discovered at every stage are listed at the bottom, with an index of the first basic block of the function in the circle. The example starts with an initial CFG in (I). Functions are first extracted, resulting in 4 functions being discovered in (II). An edge $(B8, B3)$ --- highlighted in red --- is promoted to a call in (III), as there is already a function call edge $(B1, B3)$. In (IV), $(B7, B8)$ is promoted due to the mismatch of IDs, and as a result $(B5, B8)$ is promoted as well -- both changes highlighted in red; $B8$ is added to the functions' list, as it now heads a newly discovered function. Finally, in (V), functions are re-assigned IDs, based on newly promoted edges.

\papername does not aim to recover all function boundaries present in the original application, as in some cases that would be impossible due to optimizations applied by the compiler, such as function inlining. Instead, by using its components, the tool can effectively recover the function boundaries necessary to generate a valid C or LLVM~IR. In exchange, \papername does not rely on the symbol table or debug information to recover those functions boundaries.

\subsection{Instructions Lifting}

Once the control flow is structured, the extracted functions boundaries are passed to the instruction lifting and code generation engine, alongside the original binary, to be disassembled and lifted into C. To avoid creating lifting rules for all possible hardware instructions, we first disassemble the code into TCG IR, an IR used by QEMU~\cite{qemu}, and then lift those IR operations into simple C statements. The disadvantage of indirectly lifting instructions is that complex binary instructions are translated into a sub-optimal sequence of IR operations. Nevertheless, those can often be optimized by modern compilers. However, TCG IR does not yet directly support floating-point instructions, and instead provides a framework for their emulation, significantly increasing the overhead. As a mitigation, we replace selected floating-point helpers with actual floating-point operations. However, in general, benchmarks that rely heavily on floating-point operations suffer a more significant performance degradation, after recompilation, than integer ones.

Moreover, in TCG IR, all variables representing hardware registers are considered global, and adopting such representation in C or LLVM~IR limits the optimizations that the compiler can perform. To alleviate this problem, we localize selected global variables, that we know cannot be accessed outside the function scope. This localization is possible because these variables correspond to registers in the original binary, allowing us to make additional assumptions --- such as the absence of potential side effects --- that would not hold for arbitrary global variables. This enables \emph{liveness analysis}, similar to one used for register allocation. In addition, such analysis can use additional knowledge based on the architecture ABI, about the scope of the registers.

Finally, the code generation engine takes lifted instructions and generates C code that can be subsequently recompiled. The current implementation does not map stack slots back to the local variables. Instead, the code implements a memory model that consists of two stacks: native and emulated. The native stack stores return addresses, registers spills, as well as any additional data introduced by the lifting process. The emulated stack, implemented as space allocated with malloc in memory, holds any stack data, which includes local variables, of the original binary.

\textit{Undiscovered Control Flow} --- As \papername may not lift the whole binary, this may cause the debloated application to fail. For example, if a specific input tries to execute a path that was not executed or discovered in the lifting process. We implement two mechanisms to handle it, with the user selecting one or another (but not both). The fail-safe approach guards unexplored branches with \texttt{assert} statements, so that the recompiled binary stops in a controlled manner, if an unexplored path is taken. This is a desired behavior, especially if such inputs are indicative of security attacks. A transparent approach, on the other hand, removes unexplored branches in the code, effectively promoting some of the conditional branches to unconditional jumps. This may improve the performance and size of the recompiled binary, as the compiler can remove dead code that the unexplored paths relied on. However, this mode assumes that the application input never forces the binary into unexplored paths. Should an unexplored path be taken, the program would either produce an incorrect result, hang, or fail with a segmentation fault. Such an approach could be used with comfort zones~\cite{comfort-zone}.

\section{Evaluation}
\label{sec:evaluation}

The evaluation assesses \papername using a number of benchmarks and compares it against other existing lifting tools. More specifically the paper answers the following questions:

\begin{itemize}
    \item \textbf{RQ1}: How does \papername compare to BinRec, Instrew and RetDec in terms of lifting time?
    \item \textbf{RQ2}: How does the code coverage compare between different \papername debloating strategies (\textbf{D}, \textbf{DS1}, \textbf{DS2})?
    \item \textbf{RQ3}: How does the recompilation impact the run time of debloated binaries?
    \item \textbf{RQ4}: How does debloating impact the number of gadgets and code size?
    \item \textbf{RQ5}: How does debloating impact multi-threaded applications and linked libraries?
\end{itemize}

\begin{figure*}[t]
\centering
\begin{tikzpicture}
\begin{axis}[
    ybar,
    ymin=0,
    ymax=7.25,
    ylabel={Normalized performance overhead},
    ytick={0.5,1.5,2,2.5,3,3.5,4,4.5,5,5.5,6,6.5,7},
    extra y ticks=1,
    extra y tick style={grid=major, grid style={dashed,black}},
    symbolic x coords={400.perlbench, 401.bzip2, 403.gcc, 429.mcf, 445.gobmk, 456.hmmer, 458.sjeng, 462.libquantum, 464.h264ref, 471.omnetpp, 473.astar, 483.xalancbmk, geomean},
    xtick=data,
    ymajorgrids=true,
    grid style={dotted,gray},
    x tick label style={rotate=25},
    legend style={at={(0.57,0.95)},anchor=north,legend columns=-1},
    legend image code/.code={%
        \draw[#1, draw=none] (0cm,-0.1cm) rectangle (0.6cm,0.1cm);
    }, 
    width=\textwidth,
    height=6cm,
    bar width=6,
    enlarge x limits=0.05
]
\addplot [draw = black,
    semithick,
    pattern = north west lines,
    pattern color = black
    ] coordinates {
        (400.perlbench,4.14)
        (401.bzip2,1.10)
        (403.gcc,7.06)
        (429.mcf,1.06)
        (445.gobmk,2.89)
        (456.hmmer,1.05)
        (458.sjeng,1.39)
        (462.libquantum,1.16)
        (464.h264ref,1.63)
        (471.omnetpp,1.55)
        (473.astar,1.15)
        (483.xalancbmk,2.05)
        (geomean,1.78)
    };
 \addplot [draw = black,
    semithick,
    pattern = north east lines,
    pattern color = black
    ] coordinates {
        (400.perlbench,2.27)
        (401.bzip2,1.09)
        (403.gcc,1.70)
        (429.mcf,1.06)
        (445.gobmk,1.45)
        (456.hmmer,1.05)
        (458.sjeng,1.39)
        (462.libquantum,1.16)
        (464.h264ref,1.61)
        (471.omnetpp,1.49)
        (473.astar,1.14)
        (483.xalancbmk,1.82)
        (geomean,1.40)
    };
 \addplot [draw = black,
    semithick,
    pattern = dots,
    pattern color = black
    ] coordinates {
        (400.perlbench,3.00)
        (401.bzip2,1.943)
        (403.gcc,2.98)
        (429.mcf,1.12)
        (445.gobmk,2.30)
        (456.hmmer,1.90)
        (458.sjeng,2.08)
        (462.libquantum,1.19)
        (464.h264ref,2.05)
        (471.omnetpp,2.04)
        (473.astar,1.56)
        (483.xalancbmk,2.40)
        (geomean,1.96)
	};
\legend{\papername (dynamic tracing + static lifting), \papername (dynamic tracing), Instrew}
\end{axis}
\end{tikzpicture}
\caption{Overhead of debloating binaries --- steps (A -- D) and (A) in Figure~\ref{fig:system} --- with \papername and lifting with Instrew for SPEC CPU2006 INT benchmarks normalized to the native execution time on a single core.}
\label{fig:time}
\Description{The figure is a bar graph comparing the execution time of \papername in two modes: "Dynamic Tracing Only" and "Dynamic Tracing + Static Analysis", alongside Instrew. The execution times are normalized to the native execution.}
\end{figure*}
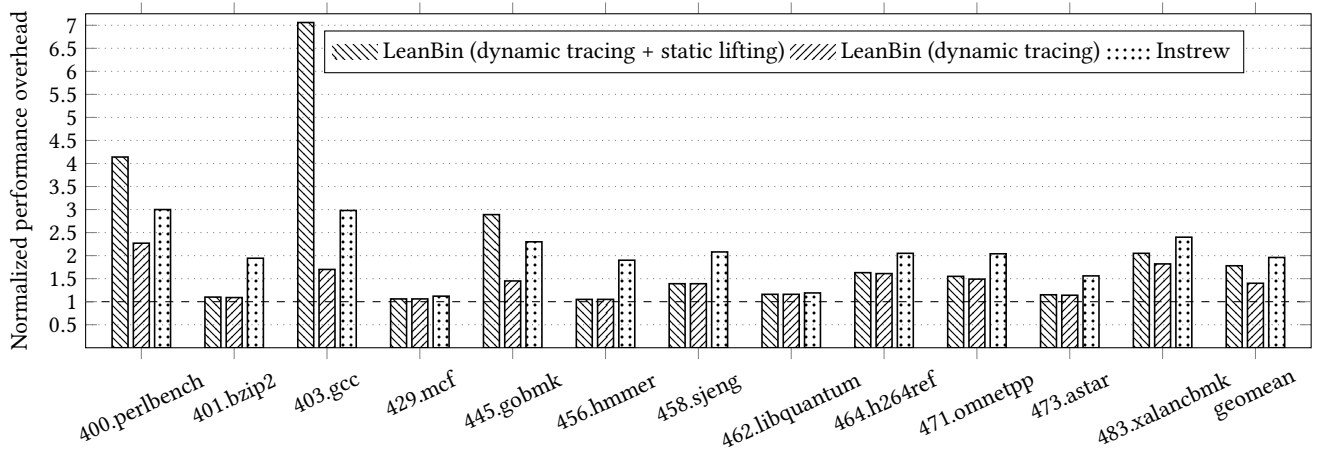

\begin{table*}[t]
    \centering
    \begin{tabular}{cccc||c||c}
    \hhline{~-----}
                        & Native [s] & \papername [s] & Instrew [s] & BinRec Intel [h]$\dag$  &RetDec (static) [s] \\ \hline
    400.perlbench (C)       & 124 & 231 & 372 & >24 & 157 \\
    401.bzip2 (C)           & 99 & 109 & 193 & 6.18     & 7 \\
    403.gcc (C)             & 38 & 270 & 114 & -    & 1159 \\
    429.mcf (C)            & 303 & 322 & 339 & 2.97      & 1 \\
    445.gobmk (C)           & 173 & 330 & 263 & 0.23    & 3 \\
    456.hmmer (C)           & 173 & 182 & 329 & -     & - \\
    458.sjeng (C)           & 392 & 546 & 816 & 13.34     & 16 \\
    462.libquantum (C)      & 216 & 250 & 258& 0.2      & 3 \\
    464.h264ref (C)         & 328 & 533 & 673 & >24     & 178 \\
    471.omnetpp (C++)        & 245 & 380 & 499 & 1.23      & 32 \\
    473.astar (C++)          & 189 & 216 & 294 & 14.93     & - \\
    483.xalancbmk (C++)       & 178 & 365 & 428 & 0.31   & 896 \\ \hline

    \rowcolor{grey} geomean & 177 & 285 & 337 & 1.68  & 31 \\ \hline
    
    \end{tabular} \\
    \medskip
    {\footnotesize  $\dag$ Execution time measured on the Intel platform with \emph{test} inputs.} \\
    \caption{Absolute single-core lifting time [seconds] of \papername, Instrew, RetDec (static) and BinRec [hours], and execution time of the native binary on SPEC CPU2006 INT benchmarks.}
    \label{table:absolute}
\end{table*}

\subsection{Experimental Setup}

\subsubsection{Hardware Setup}

The experiments run on a Gigabyte R272-P30 Ampere Altra server with 80 ARMv8.2 cores (AArch64) and 512GB of RAM. This ARM machine runs Ubuntu 20.04 LTS with kernel version 5.15.0. The evaluation of BinRec performance executes on a 10-core Intel i9-7900X machine running at 3.30 GHz with 64 GB of RAM, Ubuntu 20.04 LTS, and kernel version 5.15.0.

\subsubsection{Software Setup}

The proposed design is implemented as a plugin to a dynamic binary modification (DBM) tool MAMBO~\cite{mambo}. The plugin acts as a concrete execution tracer and performs subsequent analysis, structuring, disassembly, and lifting. MAMBO has been selected as it is a tool specifically optimized for ARM architectures, so the tracing overhead can be kept to the minimum. For disassembly into TCG IR, we modified the QEMU-based UNICORN library~\cite{unicorn} used for processor emulation. Since QEMU cannot be built as a standalone library, using UNICORN enables \papername to have access to actively maintained and updated ISA support for many architectures that can be linked with the lifter.

Since the dynamic tracer is implemented as a plugin using a set of callbacks defined by MAMBO API, it makes it possible to support x86 (or other) architectures in the future by using other major DBM tools such as PIN~\cite{pin} or DynamoRIO~\cite{dynamorio} that also provide similar APIs. Additionally, the current version of the UNICORN library already supports all common architectures, making the port to x86 primarily a (moderate) engineering effort.

The evaluation compares \papername with Instrew, RetDec, and BinRec. However, RetDec does not support debloating, and Instrew does not generate a standalone binary. For AArch64, there is no static lifter capable of executing complex benchmarks; RetDec is the most complete tool available. The direct comparison does not include BinRec as it does not support AArch64.

\subsubsection{Benchmarks}
\label{sec:benchmarks}

We run \papername, Instrew, BinRec and RetDec with default settings on a single core with SPEC CPU2006. We selected SPEC CPU2006 because it is the main benchmark that works across Instrew, BinRec, and RetDec, and it represents a wide range of non-trivial workloads with compiled binaries containing both direct and indirect branches. Despite SPEC benchmarks being primarily used for performance analysis, they are the main benchmarks used in the evaluation of binary lifters~\cite{secondwrite-1,binrec} due to the complexity and diversity of scenarios they represent.

We compiled the benchmarks with gcc 9.4.0 as position\hyp{}independent dynamically linked executables with the \texttt{-O3} optimization level and with \texttt{-fno-tree-vectorize}. The vectorization is disabled to avoid measuring the overhead of the sub-optimal disassembly of vector instructions into TCG IR. Following the evaluation of BinRec \cite{binrec}, we focus on the integer benchmarks. As TCG emulates floating-point operations, including the results for SPEC2006 FP would mainly measure the overhead of emulating floating-point with integer instructions. All experiments check for correct execution. As \papername currently does not support C++ exceptions, the \emph{omnetpp} benchmark executes and generates results correctly, but the exit is not clean due to an uncaught exception. For BinRec, we compile SPEC benchmarks for x86 (32-bits) with gcc 4.8.0, to replicate the setup in Altinay \emph{et al.}~\cite{binrec}.

To stress the multi-threaded capabilities of \papername we use the Phoenix~\cite{phoenix} benchmarks, as they were previously used in a similar context~\cite{lasagne}. The suite evaluates map-reduce applications on multi-core systems and takes advantage of all 80 cores. The compiler optimization level was set to the default \texttt{-O3}. We excluded \emph{histogram} as it encounters a segmentation fault even when running natively, and \emph{kmeans} as it enters a deadlock condition under MAMBO. BinRec does not support multi-threaded applications, and Instrew fails with the multi-threaded benchmarks due to unsupported syscalls.

\subsection{Experimental Results}

\subsubsection{RQ1: Debloating Time}
\label{sec:debloating}

The execution time is an important consideration for the design of a debloating tool, as well as lifters. When debloating is used for security, and lifting to aid in reverse engineering, faster execution improves the productivity of security experts. The existing literature acknowledges this, and performance is always a key part of the evaluation~\cite{lmcas, chisel, trimmer}.

Our binary debloating approach aims to reduce the overhead of running the binary under the tool compared to other dynamic approaches, hence reducing the overall lifting and debloating time --- in other words, \papername aims to provide faster lifting times. To evaluate this overhead, we run SPEC CPU2006 INT benchmarks for \emph{ref} inputs with and without our tool. We picked the one \emph{ref} input with the longest native execution time for each benchmark. This follows the evaluation approach of BinRec that: ``reports the maximum time among the reference workloads'' \cite{binrec}. In Figure~\ref{fig:time}, we plotted two overheads for each benchmark (expressed as a slowdown normalized to the native execution): the dynamic overhead of the control tracing, and the overhead of the entire process. We only evaluate the lifting overhead with strategy \textbf{D}, as the additional heuristic-free static analysis has a negligible ($<1s$) execution time.

The slowdown of the dynamic tracing varies between 1.05$\times$ and 2.27$\times$ with a geomean of 1.40$\times$. However, for the entire process, the slowdown can be as high as 7.06$\times$. For the majority of benchmarks, the static part has a negligible impact on the overall execution time, with an exception for \emph{gcc} and \emph{gobmk}, where at least half of the lifting time is spent on lifting basic blocks into IR. Those two benchmarks have the highest number of uniquely executed basic blocks so that the static lifting dominates the overall execution time compared to smaller binaries. In the \emph{gcc} case, a single run is relatively short (about 90 seconds) but generates a trace of substantial size (millions of executed basic blocks). As a result, most of the time is spent on statically processing this trace. However, since the static lifting performance depends only on the application size and number of lifted basic blocks, it can be considered a \emph{fixed} overhead, as it is not a function of the run time of the binary. As such, a relative comparison to the program's native execution time may appear disproportionate for large, short-running applications.

\begin{figure*}[t]
\centering
\begin{tikzpicture}
\begin{axis}[
    ybar,
    ymin=0,
    ymax=100,
    ytick={0, 25, 50, 75, 100},
    extra y tick style={grid=major, grid style={dashed,black}},
    ylabel={Code Coverage [\%]},
    symbolic x coords={400.perlbench, 401.bzip2, 403.gcc, 429.mcf, 445.gobmk, 456.hmmer, 458.sjeng, 462.libquantum, 464.h264ref, 471.omnetpp, 473.astar, 483.xalancbmk, geomean},
    xtick=data,
    ymajorgrids=true,
    grid style={dotted,gray},
    x tick label style={rotate=25},
    legend style={at={(0.875,0.95)},anchor=north,legend columns=-1},
    legend image code/.code={%
        \draw[#1, draw=none] (0cm,-0.1cm) rectangle (0.6cm,0.1cm);
    }, 
    width=\textwidth,
    height=4.5cm,
    bar width=8,
    enlarge x limits=0.05
]
\addplot [draw = black,
    semithick,
    pattern = dots,
    pattern color = black
    ] coordinates {
        (400.perlbench,0.00)
        (401.bzip2,66.20)
        (403.gcc,31.72)
        (429.mcf,73.86)
        (445.gobmk,57.69)
        (456.hmmer,9.92)
        (458.sjeng,40.99)
        (462.libquantum,32.28)
        (464.h264ref,33.77)
        (471.omnetpp,25.90)
        (473.astar,67.27) 
        (483.xalancbmk,16.07)
        (geomean,35.45)
	};
 \addplot [draw = black,
    semithick,
    pattern = north west lines,
    pattern color = black
    ] coordinates {
        (400.perlbench,47.59)
        (401.bzip2,81.68)
        (403.gcc,46.05)
        (429.mcf,81.89)
        (445.gobmk,64.63)
        (456.hmmer,11.68)
        (458.sjeng,49.51)
        (462.libquantum,41.16)
        (464.h264ref,48.37)
        (471.omnetpp,29.74)
        (473.astar,69.45) 
        (483.xalancbmk,20.20)
        (geomean,35.45)
	};
  \addplot [draw = black,
    semithick,
    pattern = north east lines,
    pattern color = black
    ] coordinates {
        (400.perlbench,65.01)
        (401.bzip2,89.56)
        (403.gcc,77.63)
        (429.mcf,88.55)
        (445.gobmk,76.17)
        (456.hmmer,44.42)
        (458.sjeng,93.77)
        (462.libquantum,56.46)
        (464.h264ref,89.01)
        (471.omnetpp,41.59)
        (473.astar,71.40) 
        (483.xalancbmk,26.34)
        (geomean,64.30)
	};
\legend{\textbf{D}, \textbf{DS1}, \textbf{DS2}}
\end{axis}
\end{tikzpicture}
\caption{Coverage of SPEC CPU2006 INT binaries debloated with \papername utilizing D, DS1 and DS2 strategies.}
\label{fig:blocks}
\Description{The figure is a bar graph displaying the code coverage of evaluated SPEC CPU2006 benchmarks using three different debloating strategies: "D," "DS1," and "DS2".}
\end{figure*}
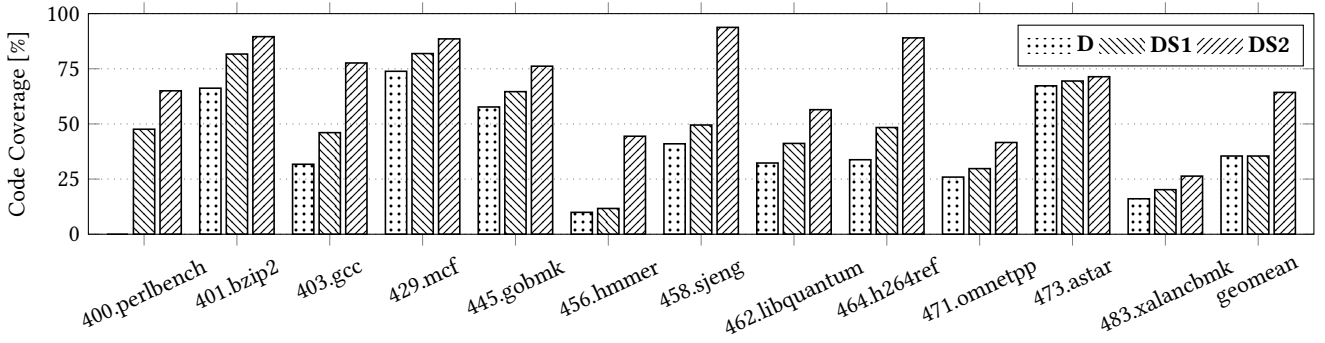

With dynamic control flow tracing, there are three factors impacting the execution time: (a) inherent overhead of the tracing tool; (b) overhead introduced by instrumenting basic blocks via callbacks; and (c) overhead of running code with instrumentation. To minimize the first factor, we use a low-overhead DBM tool, MAMBO. The second factor depends on the number of unique executed basic blocks. Each new basic block has to be scanned, recorded, and instrumented if necessary. Finally, executing code with instrumentation increases the overall instruction count, and slows down the application. We minimize this overhead by instrumenting only indirect branches, which are less prevalent, and use a lightweight hash-map-based instrumentation written in assembly.

To better quantify gains from using our lifter, we compare \papername with Instrew, BinRec, and RetDec. All results were presented in Figure~\ref{fig:time} and Table~\ref{table:absolute}. As BinRec does not currently support AArch64 and was tested with smaller inputs, we do not aim to provide a definite comparison with it, but rather to show that even for small \emph{test} inputs, resulting in a native execution of few seconds without any tool, BinRec performs significantly slower than \papername and Instrew running with \emph{ref} inputs.

Although Instrew does not enable lifting and recompilation into a standalone binary and only generates LLVM~IR on the fly, it is currently the closest project that our tool can be compared against. It generates LLVM~IR and can lift observed execution. For all benchmarks, \papername features a lower dynamic overhead, as it does not have to perform any lifting during run time. However, for benchmarks with a very large number of lifted basic blocks, (e.g., \emph{gcc}), \papername incurs a large static lifting time, resulting in a higher overall relative overhead compared to Instrew.

This shows that we offer a competitive performance over Instrew despite offering more functionality; full lifting. Moreover, once the binary code is lifted and debloated, subsequent runs of the new binary do not require a tracer, thus lowering the execution overhead, something Instrew does by non-persistent caching of the lifted code.

Finally, for completeness, we present results for a static lifter, RetDec. In most cases, RetDec outperforms dynamic approaches, but it runs slower with the most complex binaries: \emph{gcc} and \emph{xalancbmk}. Note that RetDec does not guarantee that the lifted code is correct, and it cannot be debloated. It also fails to lift 2 tested benchmarks.

\begin{summary}
The normalized lifting time is between $1.05\times$ and $2.27\times$ with a geomean of $1.40\times$ for dynamic tracing, and between $1.05\times$ and $7.06\times$ and geomean of $1.78\times$ for full lifting, outperforming both BinRec and Instrew in most of the cases.
\end{summary}

\subsubsection{RQ2: Code Coverage}
\label{sec:coverage}

\papername relies on concrete inputs and heuristic-free static analysis to recover the control flow, however complete coverage is not a goal. The value of \papername, however, lies elsewhere, as it is a binary debloating tool, and enables a number of tasks such as: creating hardened binaries that can execute more securely for a set of inputs, or trace analysis allowing to ensure that the program does not exhibit insecure behavior on a given path.

The evaluation focuses solely on the coverage of \papername, as the coverage of Instrew is no better than the one of Leanbin with static analysis disabled. For RetDec, coverage is assumed to be 100\% when the lifting process is successful. The comparison with BinRec is more nuanced. While the concrete dynamic coverage of Leanbin and BinRec is identical, BinRec utilizes concolic execution to explore additional branches dynamically, extending beyond the concrete input. However, BinRec's coverage depends on how long the lifter is allowed to run, potentially reaching 100\% if given sufficient time. As such the direct coverage comparison is non-trivial.

We measure the coverage of binaries for given references inputs with strategies\textbf{D}, \textbf{DS1} and \textbf{DS2}. The coverage is measured as a percentage of the instructions lifted from the \texttt{.text} section of the original binary. The results are presented in Figure~\ref{fig:blocks}.

More irregular and complex applications that have to account for many different cases (e.g., \emph{gcc}, \emph{xalancbmk}) exhibit a low code coverage when only dynamic analysis is used, whereas more data-oriented applications that apply similar operations to all the data, e.g., \emph{bzip2}, \emph{mcf}, have a much higher dynamic coverage. At the same time, adding static analysis can provide significant benefits, when high coverage is required, with some applications reaching 90\% or more. It is also important to note that 100\% is not achievable using the current metric, as some code is never lifted, e.g., the \texttt{\_start} function or any dead code. \emph{perlbench} is excluded in strategy \textbf{D}, as lifting it with a single trace excessively restricts the new binary, causing failures due to changes in memory alignments.

Finally, this paper presents a number of strategies, but the analysis provided can be expanded in the future to suit specific applications and required guarantees. The proposed approach shows that combining dynamic and static analysis is a promising approach when code coverage, lifting correctness, and execution time have to be balanced. As such, \papername is the first hybrid lifter combining static and dynamic analysis.

\begin{summary}
Code coverage is between 9.92\% and 73.86\% (geomean of 35.45\%) for strategy \textbf{D}, between 11.68\% and 81.89\% (geomean of 34.45\%) for \textbf{DS1}, and between 26.34\% and 89.56\% (geomean of 64.30\%) for \textbf{DS2}.
\end{summary}

\subsubsection{RQ3: Execution Time of Debloated Binaries}
\label{sec:recompile}

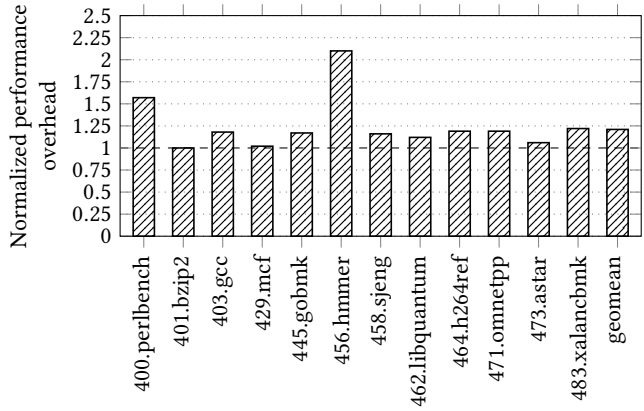
\begin{figure}[t]
\centering
\begin{tikzpicture}
\begin{axis}[
    ybar,
    ymin=0,
    ymax=2.5,
    ytick={0, 0.25, 0.5, 0.75, 1.25, 1.5, 1.75, 2, 2.25, 2.5},
    extra y ticks=1,
    extra y tick style={grid=major, grid style={dashed,black}},
    ylabel style={align=center},
    ylabel={Normalized performance\\overhead},
    symbolic x coords={400.perlbench, 401.bzip2, 403.gcc, 429.mcf, 445.gobmk, 456.hmmer, 458.sjeng, 462.libquantum, 464.h264ref, 471.omnetpp, 473.astar, 483.xalancbmk, geomean},
    xtick=data,
    ymajorgrids=true,
    grid style={dotted,gray},
    x tick label style={rotate=90},
    legend style={at={(0.5,-0.5)},anchor=north,legend columns=-1},
    width=\columnwidth,
    height=4.5cm,
    bar width=8,
    enlarge x limits=0.05
]
\addplot [draw = black,
    semithick,
    pattern = north east lines,
    pattern color = black
    ] coordinates {
        (400.perlbench,1.57)
        (401.bzip2,1.00)
        (403.gcc,1.18)
        (429.mcf,1.02)
        (445.gobmk,1.17)
        (456.hmmer,2.10)
        (458.sjeng,1.16)
        (462.libquantum,1.12)
        (464.h264ref,1.19)
        (471.omnetpp,1.19)
        (473.astar,1.06) 
        (483.xalancbmk,1.22)
        (geomean,1.21)
	};
\end{axis}
\end{tikzpicture}
\caption{Execution time of debloated binaries (\textbf{DS2} strategy) normalized to the native execution of SPEC CPU2006 INT benchmarks.}
\label{fig:performance}
\Description{The figure is a bar graph showing the execution time of evaluated SPEC CPU2006 benchmarks, normalized to the native execution.}
\end{figure}

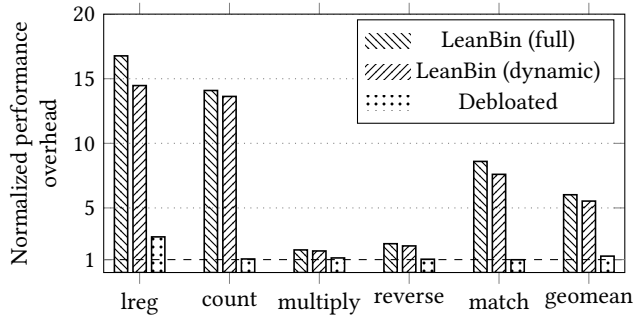
\begin{figure}[t]
\centering
\begin{tikzpicture}
\begin{axis}[
    ybar,
    ymin=0,
    ymax=20,
    ylabel style={align=center},
    ylabel={Normalized performance\\overhead},
    ytick={5,10,15,20},
    extra y ticks=1,
    extra y tick style={grid=major, grid style={dashed,black}},
    symbolic x coords={lreg, count, multiply, reverse, match, geomean},
    xtick=data,
    ymajorgrids=true,
    grid style={dotted,gray},
    x tick label style={rotate=0},
    legend image code/.code={%
        \draw[#1, draw=none] (0cm,-0.1cm) rectangle (0.6cm,0.1cm);
    }, 
    height=5cm,
    width=\columnwidth,
    bar width=5,
    enlarge x limits=0.08
]
\addplot [draw = black,
    semithick,
    pattern = north west lines,
    pattern color = black
    ] coordinates {
        (lreg,16.77)
        (multiply,1.75)
        (reverse,2.23)
        (match,8.60)
        (count,14.09)
        (geomean,6.02)
    };
 \addplot [draw = black,
    semithick,
    pattern = north east lines,
    pattern color = black
    ] coordinates {
        (lreg,14.47)
        (multiply,1.67)
        (reverse,2.06)
        (match,7.60)
        (count,13.63)
        (geomean,5.53)
    };
 \addplot [draw = black,
    semithick,
    pattern = dots,
    pattern color = black
    ] coordinates {
        (lreg,2.76)
        (multiply,1.13)
        (reverse,1.03)
        (match,0.99)
        (count,1.05)
        (geomean,1.27)
	};
\legend{\papername (full), \papername (dynamic), Debloated}
\end{axis}
\end{tikzpicture}
\caption{Overhead of debloating --- steps (A -- D) and (A) in Figure~\ref{fig:system} --- Phoenix benchmarks with \papername and performance of debloated binary normalized to the native execution running on 80 cores.}
\label{fig:phoenix}
\Description{The figure is a bar graph depicting the lifting time of five Phoenix benchmarks: "lreg", "count", "multiple", "reverse", and "match". It shows both the full lifting time and the time for dynamic tracing only. Additionally, the execution time of debloated binaries is presented. All results are normalized to the native execution.}
\end{figure}

Figure~\ref{fig:performance} assesses the performance after debloating and recompiling the lifted code with \texttt{clang} and \texttt{-O3}. We only evaluate the \textbf{DS2} strategy as the largest amount of code is generated, reducing optimization opportunities and providing the most pessimistic run time. The geomean slowdown compared to the native execution without lifter is around 1.21$\times$. The reported geomean slowdown in Altinay \emph{et al.}~\cite{binrec} for BinRec is similar to that of \papername. Benchmarks relying on floating-point helpers (e.g., \emph{hmmer}) have a higher overhead compared to mainly integer only benchmarks (e.g., \emph{bzip2}, \emph{mcf}) that achieve native or near-native performance ($1.00-1.02\times$). 

The major factor impacting the performance is the inefficiency of mapping binary into TCG IR. Using the IR simplifies the process, as only a limited number of TCG operations have to be translated into C. However, complex binary instructions are turned into a sequence of TCG operations (e.g., integer comparison) or helper functions (e.g., floating-point emulation) that cannot always be recompiled back into the original code.

\begin{summary}
The normalized execution time of recompiled binaries varies between $1.00\times$ and $2.10\times$ with a geomean of $1.21\times$.
\end{summary}

\subsubsection{RQ4: Attack Surface and Code Size Reduction}
\label{sec:gadgets}

\begin{table*}[tb]
  \centering
  \begin{tabular}{c|c|c|c|c|c|c|c|c|c|}
    \hhline{~---------}
    & \multicolumn{3}{c|}{\textbf{Debloated}} & \multicolumn{3}{c|}{\textbf{Hardened}} & \multicolumn{3}{c|}{\textbf{Size}} \\
    \hhline{~---------}
    & \textbf{D} & \textbf{DS1} & \textbf{DS2} & \textbf{D} & \textbf{DS1} & \textbf{DS2} & \textbf{D} & \textbf{DS1} & \textbf{DS2} \\
    \hline
        \emph{400.perlbench}    & -       & \color{ForestGreen}30.39\% & \color{ForestGreen}28.52\% & -       & \color{ForestGreen}22.03\% & \color{ForestGreen}22.13\% & -      & \color{ForestGreen}80.32\% & \color{OrangeRed}119.87\% \\
        \emph{401.bzip2}        & \color{ForestGreen}40.30\% & \color{ForestGreen}46.12\% & \color{ForestGreen}50.43\% & \color{ForestGreen}35.13\% & \color{ForestGreen}36.42\% & \color{ForestGreen}41.59\% & \color{ForestGreen}74.61\% & \color{OrangeRed}107.77\% & \color{OrangeRed}122.29\% \\
        \emph{403.gcc}          & \color{ForestGreen}16.43\% & \color{ForestGreen}35.96\%  & \color{ForestGreen}34.39\% & \color{ForestGreen}10.56\% & \color{ForestGreen}18.94\% & \color{ForestGreen}21.27\% & \color{ForestGreen}40.65\% & \color{ForestGreen}68.92\% & \color{OrangeRed}124.29\% \\
        \emph{429.mcf}          & \color{ForestGreen}70.54\% & \color{ForestGreen}75.52\%  & \color{ForestGreen}78.84\% & \color{ForestGreen}59.75\% & \color{ForestGreen}61.00\% & \color{ForestGreen}62.66\% & \color{ForestGreen}94.83\% & \color{OrangeRed}111.88\% & \color{OrangeRed}120.12\% \\
        \emph{445.gobmk}        & \color{ForestGreen}47.55\% & \color{ForestGreen}40.58\% & \color{ForestGreen}37.05\% & \color{ForestGreen}33.65\% & \color{ForestGreen}20.28\% & \color{ForestGreen}20.89\% & \color{ForestGreen}84.92\% & \color{OrangeRed}101.34\% & \color{OrangeRed}123.51\% \\
        \emph{456.hmmer}        & \color{ForestGreen}11.26\% & \color{ForestGreen}15.41\%  & \color{ForestGreen}19.98\% & \color{ForestGreen}8.87\% & \color{ForestGreen}9.88\% & \color{ForestGreen}11.96\% & \color{ForestGreen}18.82\% & \color{ForestGreen}25.90\% & \color{ForestGreen}86.70\% \\
        \emph{458.sjeng}        & \color{ForestGreen}52.76\% & \color{ForestGreen}54.01\%  & \color{ForestGreen}68.67\% & \color{ForestGreen}37.22\% & \color{ForestGreen}39.10\% & \color{ForestGreen}51.00\% & \color{ForestGreen}75.44\% & \color{ForestGreen}59.20\% & \color{OrangeRed}123.59\% \\
        \emph{462.libquantum}   & \color{ForestGreen}37.55\% & \color{ForestGreen}38.80\% & \color{ForestGreen}42.74\% & \color{ForestGreen}32.78\% & \color{ForestGreen}29.05\% & \color{ForestGreen}32.57\% & \color{ForestGreen}52.50\% & \color{ForestGreen}66.05\% & \color{OrangeRed}105.31\% \\
        \emph{464.h264ref}      & \color{ForestGreen}17.34\% & \color{ForestGreen}22.29\% & \color{ForestGreen}26.87\% & \color{ForestGreen}13.45\% & \color{ForestGreen}15.95\% & \color{ForestGreen}20.09\% & \color{ForestGreen}41.01\% & \color{ForestGreen}68.98\% & \color{OrangeRed}131.16\% \\
        \emph{471.omnetpp}      & \color{ForestGreen}4.96\% & \color{ForestGreen}7.18\% & \color{ForestGreen}6.81\% & \color{ForestGreen}2.95\% & \color{ForestGreen}3.94\% & \color{ForestGreen}4.15\% & \color{ForestGreen}38.88\% & \color{ForestGreen}68.01\% & \color{ForestGreen}76.17\% \\
        \emph{473.astar}        & \color{ForestGreen}55.10\% & \color{ForestGreen}57.37\% & \color{ForestGreen}57.37\% & \color{ForestGreen}40.36\% & \color{ForestGreen}41.50\% & \color{ForestGreen}40.36\% & \color{OrangeRed}102.86\% & \color{OrangeRed}108.97\% & \color{OrangeRed}113.18\% \\
        \emph{483.xalancbmk}    & \color{ForestGreen}6.79\% & \color{ForestGreen}7.64\% & \color{ForestGreen}6.53\% & \color{ForestGreen}2.12\% & \color{ForestGreen}4.81\% & \color{ForestGreen}4.29\% & \color{ForestGreen}33.42\% & \color{ForestGreen}29.16\% & \color{ForestGreen}42.19\% \\
        \hline
        \rowcolor{grey} \emph{geomean} & 24.10\% & 28.97\% & 30.22\% & 16.65\% & 19.28\% & 21.00\% & 53.59\% & 68.25\% & 103.17\% \\ 
    \hline
  \end{tabular}
  \caption{Reduction in the number of unique gadgets expressed as the percentage of gadgets of the original binary and code size expressed as the percentage of the \texttt{.text} section size of the original binary for SPEC CPU2006 INT benchmarks.}
  \label{table:gadgets}
\end{table*}

In this section, we evaluate the ability of \papername to create specialized and hardened binaries that can be used to run for a set of defined inputs.

\textit{Attack Surface Reduction} ---
The attack surface is a sum of all the ways (e.g., return-oriented programming attacks, side channel attacks) a binary can be attacked or compromised. To quantify the increase in the security of the application, we use Ropper~\cite{ropper}, a tool for counting gadgets in a binary, and present results in Table~\ref{table:gadgets}. \textit{Gadgets} are sequences of instructions that can be chained by a malicious actor in order to gain control over the system. Thus, a specialized binary with a smaller number of gadgets is more secure against return-oriented programming attacks; and as a result the attack surface is reduced.

We debloated SPEC CPU2006 INT benchmarks with selected \emph{ref} inputs (the same as previously) and recompiled the code, creating specialized input-dependent binaries using the three proposed strategies. The number of gadgets expressed as the percentage of gadgets of the original binary varies between 4.96\% and 78.84\%, with a geomean of 24.10\% to 30.22\%. Unsurprisingly, the largest benchmarks see the biggest reduction in the number of gadgets, as they cover many different conditions for various inputs. The smallest benchmark, \emph{mcf}, with the highest code coverage results in the smallest reduction.

Since the lifted code can be recompiled with any tool, we can take further advantage of AArch64 architectural features to improve the security of the code. Since ARMv8.3, there is support for Pointer Authentication Code (PAC). Modern compilers can take advantage of this feature, for example, to check that return addresses from functions have not been modified. To quantify the effect of PAC, we compiled the lifted code with PAC enabled to further reduce the number of gadgets. For many benchmarks, the extra reduction was relatively modest but comes at zero cost to developers. However, for some of the benchmarks the remaining number of gadgets was halved, bringing the total reduced number of gadgets for \emph{xalancmnk} (strategy \textbf{D}) to 2.12\%.

\textit{Code Size Reduction} ---
Finally, we measure the code size (\texttt{.text} section size) of specialized binaries recompiled with \texttt{-Os}. The results are presented in Table \ref{table:gadgets}. The code reduction, expressed as the percentage of the size of the original binary, can be as low as 18.82\% with geomean of 53.59\% for \textbf{D}, and 42.19\% with geomean of 103.17\% for \textbf{DS2}. It shows that debloated binaries not only have fewer gadgets but in many cases feature a smaller code size. \emph{astar} does not produce smaller code with any strategy due to the combination of a 73\% coverage and TCG helpers called in 71 instances limiting compiler optimizations. In the case of \textbf{DS2} the high coverage compounded with TCG inefficiencies results in larger code size for many benchmarks, however future improvements to the disassembler can mitigate this issue.

\begin{summary}
The percentage (geomean) of gadgets of the original binary is 24.10\%, 28.97\% and 30.22\% and the percentage of the code size is 53.59\%, 68.25\% and 103.17\% for \textbf{D}, \textbf{DS1} and \textbf{DS2} respectively.
\end{summary}

\subsubsection{RQ5: Additional Use Cases}
\label{sec:multithreaded}
\label{sec:sqlite}
Finally, we evaluate the capability of \papername in lifting multi-threaded applications, and binaries alongside their dependencies.

\textit{Multi-threaded Applications} ---
Compared to BinRec (no support for multi-threaded applications) and Instrew (no debloating) \papername can lift multi-threaded applications. To showcase this ability, we lift, debloat, and run 6 benchmarks from the Phoenix~\cite{phoenix} benchmark suite, and present results in Figure~\ref{fig:phoenix}. The benchmark \emph{pca} is not included in the graph due to the short (<10ms) native execution time, which generates a disproportionate lifting overhead (>$1000\times$).

Compared to SPEC benchmarks, the lifting overhead is higher due to Phoenix benchmarks executing on 80 cores for a short period of time. As the compute time on each thread is relatively low, the DBM overhead dominates the overall execution time. Nevertheless, \papername correctly lifts all the benchmarks. The overhead after recompilation is comparable to the SPEC CPU2006 INT benchmarks, with \emph{linear regression} being an outlier due to its reliance on a long multiplication which is implemented as TCG helpers.

\begin{table}[t]
  \centering
  \begin{tabular}{p{1.5cm}p{1.8cm}p{1.8cm}p{1.8cm}}
    \hhline{~---}
    & \textbf{Lib (O0)} & \textbf{Lib (O3)} & \textbf{Full} \\
    \hline
        \textbf{User (O0)} & 1.24$\times$ (6.30\%) & 1.01$\times$ (4.30\%) & - \\
        \textbf{User (O3)} & 1.23$\times$ (6.50\%) & 1.06$\times$(4.47\%) & - \\
        \textbf{Full} & - & - & 0.98$\times$ (3.65\%) \\
    \hline
  \end{tabular}
  \caption{Execution time speed-up ($\times$) over the original binary and the number of gadgets expressed as the percentage of gadgets of the original binary (\%) for SQLite use case.}
  \label{table:sqlite-results}
\end{table}

\textit{Debloating Including Linked Binaries} ---
\papername can be used to debloat not only application binaries but also their dependencies, creating a more secure, faster, and specialized binary. We use SQLite as a use case by building a database user application (an interactive shell) dynamically linked against the sqlite3 library. SQLite has been selected as SQL applications are widely used and familiar to many readers. Our approach is however designed to allow the debloating of any libraries linked against the user binary. We then run the application with a sample database~\footnote{\url{https://github.com/lerocha/chinook-database}} and a compute-heavy \texttt{CROSS JOIN} query. As a result, the new binary only contains code that was previously executed from both the binary and the library. Since all the code now resides in a single file, the compiler is able to perform additional optimizations between the user and library code.

We ran the experiments in 5 different configurations: both SQL shell and library un-optimized; one of the components optimized; both components optimized. We also compiled together shell and library code, allowing the compiler to perform optimizations across both modules. After the debloating, the percentage of the number of gadgets of the original binary and the SQLite library is between 3.65\% and 6.50\%. Moreover, by specializing the binary for selected inputs we can get a performance improvement (1.01$\times$ to 1.24$\times$), showing that the debloater can be used as a post-release cross-library optimizer, even for initially optimized programs, by combining user and library code. The only case where a 1.02$\times$ slowdown is seen is for the fully optimized single binary, as the original compiler had already performed cross-module optimizations.

\begin{summary}
\papername can debloat both multi-threaded applications and user binaries alongside their dependencies.
\end{summary}

\section{Related Work}
\label{sec:rw}

\subsection{Software Debloating}

Table \ref{table:debloating-tools} summarizes the main related work on program debloating and specialization. The next paragraphs explain the related work, moving column by column from left to right.

\begin{table*}[t]
    \centering
    \begin{tabular}{p{2.9cm} p{0.8cm} p{1.4cm} p{1.2cm} p{2.0cm} p{3.0cm} p{1.7cm} p{1.0cm}}
        \hhline{~-------}
        & \textbf{Type} & \textbf{Source} & \textbf{Scope} & \textbf{Granularity} & \textbf{Policy} & \textbf{Overhead} & \textbf{Year} \\
        \hline

        Squeeze~\cite{squeeze} & D, S & Binary & P, L & Basic Blocks & Analysis + Traces & No & 2000 \\

        Squeeze++~\cite{squeezepp} & D, S & Binary & P, L & Basic Blocks & Analysis + Traces & No & 2005 \\
        DIABLO~\cite{diablo-1, diablo-2, diablo-3} & D, S & Binary & P, L, K & Basic Blocks & Analysis + Traces & No & 2005 \\ 
        
        OCCAM~\cite{occam} & S & LLVM~IR & P & Basic Blocks & Analysis & No & 2015 \\ 
        
        Feature~\cite{feature-based} & S & Bytecode & P & Functions & Analysis & No & 2016 \\ 
        JRed~\cite{jred} & S & Bytecode & P, L & Functions & Analysis & No & 2016 \\ 

        Shredder~\cite{shredder} & D, S & Binary & P & Call Sites & Analysis & Run-time & 2018 \\ 
        TOSS~\cite{toss} & D & Binary & P & Basic Blocks & Config +~Traces & Debloating & 2018 \\ 
        Piece-wise~\cite{piece-wise} & D, S & LLVM~IR$\dag$ & L & Functions & Analysis & Load-time & 2018 \\ 
        CHISEL~\cite{chisel} & S & C/C++ & P & Basic Blocks & Analysis + Config & Debloating & 2018 \\ 
        Perses~\cite{perses} & S & C/C++ & P & Basic Blocks & Analysis & No & 2018 \\ 
        BinRec~\cite{binrec-debloat} & D & Binary & P & Basic Blocks & Traces & Debloating & 2018 \\ 

        RAZOR~\cite{razor} & D, S & Binary & P & Basic Blocks & Analysis +~Traces & Debloating & 2019 \\ 
        Shrinking~\cite{honey} & D, S & Binary & P, L & Functions & Analysis + Traces & Debloating & 2019 \\ 
        BinTrimmer~\cite{bintrimmer} & S & Binary & P & Basic Blocks & Analysis & No & 2019 \\ 
        ASSS~\cite{asss} & S & LLVM~IR & P, L & Functions & Analysis & No & 2019 \\ 
        
        BlankIt~\cite{blankit} & D & Binary & L & Functions & Analysis & Run-time & 2020 \\ 
        Nibbler~\cite{nibbler} & S & Binary & P, L & Functions & Analysis & No & 2020 \\ 
        DomGad~\cite{domgad} & S & C/C++ & P & Basic Blocks & Analysis & No & 2020 \\ 
        Guided \mbox{Linking}~\cite{guided-linking} & S & LLVM~IR & P, L & Functions & Analysis & No & 2020 \\ 

        ANCILE~\cite{ancile} & D & LLVM~IR & P & Basic Blocks & Traces & Debloating & 2021 \\ 
        LMCAS~\cite{lmcas} & S & LLVM~IR & P & Basic Blocks & Analysis + Config & No & 2021 \\ 
        TRIMMER~\cite{trimmer} & S & LLVM~IR & P & Basic Blocks & Analysis + Config & No & 2021 \\ 

        uTrimmer~\cite{utrimmer} & S & Binary & P, L & Basic Blocks & Analysis & No & 2022 \\ 
        
        TLASR~\cite{tlasr} & D & Binary & P, L & Functions & Analysis & Run-time & 2023 \\ 
        OCCAMv2~\cite{occamv2} & D, S & LLVM~IR & P, L & Basic Blocks & Analysis & No & 2023 \\ 

        \textbf{\papername} & \textbf{D, S} & \textbf{Binary} & \textbf{P, L} & \textbf{Basic Blocks} & \textbf{Analysis + Traces} & \textbf{Debloating} & \textbf{2024} \\

    \hline
    \end{tabular}\\
    
    \medskip
    {\footnotesize Type: D - Dynamic, S - Static.}   {\footnotesize Scope: P - Program, L - Library, K - Kernel.} 
    {\footnotesize $\dag$ Only shared libraries source code required.} \\
    \caption{Comparison of different software debloating tools.}
    \label{table:debloating-tools}

    \medskip

    \centering
    \begin{tabular}{p{3.0cm} p{2.2cm} p{2.0cm} p{2.5cm} p{1.5cm} p{1.5cm} P{1.5cm}}
        \hhline{~------}
        & \mbox{\textbf{Static Lifters$*$}} & \textbf{HQEMU}~\cite{hqemu} & \textbf{Instrew}~\cite{instrew-2} & \textbf{BinRec}~\cite{binrec} & \textbf{TOP}~\cite{top} & \textbf{\papername}\\
        \hline
        CFG \mbox{Recovery} & Static & Dynamic & Dynamic & Dynamic & Dynamic & Hybrid\\

        Code Coverage & Complete & Trace & Trace & Varies & Trace & Varies\\

        Instructions Lifting & Static & Dynamic & Dynamic & Dynamic & Dynamic & Static\\

        Debloating & No & No & No & Yes & No & Yes\\
        
        Execution & - & QEMU & Native & S\textsuperscript{2}E & QEMU & Native\\

        Generates \mbox{Binary} & Yes & No & No & Yes & Yes & Yes\\

        Architectures & Many & Many & x86, AArch64, RV64 & x86 & x86 & AArch64\\

        \mbox{Multi-Threaded Binaries} & - & Yes & Yes & No & Yes & Yes\\

        Run Time Overhead & - & Low & Low & High & N/A & Low\\
    \hline
    \end{tabular}\\
    
    \medskip
    {\footnotesize  $*$ Static Lifters: McSema~\cite{mcsema}, reopt~\cite{reopt}, SecondWrite~\cite{secondwrite-1, secondwrite-2}, mctoll~\cite{mctoll}, LLBT~\cite{llbt}, RevGen~\cite{revgen}, rev.ng~\cite{revng}, RetDec~\cite{retdec}, FoxDec~\cite{foxdec-2}}\\
    \caption{Comparison of different binary lifting tools.}
    \label{table:comparison}
\end{table*}

\textit{Type} --- Debloating can be done either based on static \cite{trimmer, utrimmer} or dynamic \cite{binrec-debloat, ancile} information, or the combination of both ~\cite{honey, occamv2}. The majority of dynamic and hybrid tools suffer from the run-time overhead that is experienced only once when the software is debloated and subsequent runs are free from the overhead. However, three solutions~\cite{piece-wise, blankit, tlasr} require additional runtime, resulting in overhead for every execution. The \papername tool, as well as BinRec~\cite{binrec-debloat}, the closest related tool, both suffer from the dynamic overhead. However, \papername's overhead is lower compared to BinRec.

\textit{Source} --- Debloating can be done either at the source code \cite{trimmer, occamv2} or binary level \cite{nibbler, utrimmer}. Source code debloating simplifies analysis and transformations, compared to tools targeting binaries, however, it cannot be applied to production binaries. Both \papername and BinRec~\cite{binrec-debloat} combine a wider applicability of binary approaches with an ability to reuse existing source code analysis and transformations.

\textit{Scope} --- Debloating can be done at the user application level, libraries or kernel level, or combination of any of them. The only tool targeting the kernel space is DIABLO~\cite{diablo-3}, with remaining tools targeting either user application \cite{binrec-debloat, ancile, trimmer}, libraries \cite{piece-wise, blankit} or both \cite{utrimmer, occamv2}. \papername enables creating specialized binaries that combine both the user binary and its dependencies.

\textit{Granularity} --- Debloating can either remove whole functions \cite{guided-linking, tlasr} or individual basic blocks \cite{binrec-debloat, trimmer, utrimmer}. Shredder~\cite{shredder}, an exception, specializes function arguments at the call site. \papername enables fine-grained debloating at the basic block level.

\subsection{Binary Lifting}

Table \ref{table:comparison} summarizes the related work on binary lifting, whereas Hazelwood \cite{dbm-lecture} and Wenzi \textit{et al.} \cite{binary-rewriting-survey} provide context on dynamic binary modification and binary rewriting. The next paragraphs explain the related work, moving columns by column.

\textit{Static Lifting} ---
The majority of lifters (as listed in Table~\ref{table:comparison}) use purely static analyses, with their limitations previously highlighted in Altinay \emph{et al.}\cite{binrec}. All of them, but RetDec \cite{retdec} and FoxDec \cite{foxdec-2} that target C, lift binaries into LLVM IR. Whereas RetDec and FoxDec focus on generating high-quality C code, \papername generates C code that be recompiled to generate debloated binaries and LLVM IR, avoiding dependencies on a specific IR version.

Compared to static lifters, \papername precisely lifts observed execution traces, and uses heuristic-free static analysis, at the potential expense of a lower code coverage. For ARM, no existing static lifter can currently lift SPEC CPU benchmarks. RecDec represents the best-case scenario, but even then, it does not support recompilation. For x86, even popular static lifters, such as McSema \cite{mcsema} and rev.ng \cite{revng}, fail to lift and recompile SPEC CPU 2006 benchmarks \cite{binrec}. 

\textit{Lifting for Dynamic Binary Instrumentation} ---
A number of dynamic binary instrumentation (DBI) and translation (DBT) tools internally lift binary instructions to LLVM IR. HQEMU \cite{hqemu} and DBILL \cite{dbill} dynamically lift TCG IR into LLVM IR, to enable JIT compilation of the emulated binary inside QEMU \cite{qemu}. Instrew \cite{instrew-2}, generates LLVM IR as a part of their instrumentation and translation process without QEMU. It translates native instructions directly into LLVM IR, improving the overall performance and reducing the runtime overhead. Direct comparison with HQEMU is not included, as Instrew is on average faster, making the extra comparison redundant.

Neither of the approaches generates code that can be compiled into a standalone binary, and they only use LLVM IR internally, so changes to the code are not persistent. Moreover, the task of lifting within DBI/DBT is simpler compared to the one of \papername. The control-flow recovery is more localized (superblock granularity), and the whole control-flow graph is not considered. Instrew follows only direct branches and stops lifting on function calls, returns, and indirect branches. It then releases control back to DBI to dispatch the next basic block and start the lifting process again. Our approach generates a single coherent program that can be run without a dynamic dispatcher. Finally, the coverage of \papername is no worse than the one of HQEMU and Instrew, as the latter two tools lift executed traces, but do not use static analysis.

\textit{Dynamic Lifting} ---
The only current state-of-the-art dynamic lifter, BinRec \cite{binrec}, attempts to achieve precise and correct lifting with an ability to achieve higher code coverage. It uses the S\textsuperscript{2}E \cite{s2e} platform to execute the binary and collect execution traces, either using concrete or concolic execution. Such an approach suffers from exceedingly high lifting overhead, especially when the concolic execution is used to maximize the coverage.

On the other hand, \papername dynamically collects observed execution traces, and augments such dynamically collected CFG with heuristic-free static analysis to further expand the coverage - a novel hybrid scheme not used by any other lifter. Our tool achieves high performance by executing the binary natively under a lightweight tracer, as opposed to using S\textsuperscript{2}E, and defers the actual lifting of instructions and additional analysis until the execution is finished. Moreover, \papername provides basic support for multi-threaded applications, something not available with BinRec.

\textit{Lifting of Observed Execution Traces} ---
TOP~\cite{top} (trace-oriented programming) proposes lifting observed execution traces into C with embedded assembly. If static analysis is not considered, conceptually, this approach is similar to \papername, however, the generated code is more restricted. TOP relies heavily on the inlined assembly to represent the body of basic blocks and only uses C to represent control flow and function interfaces. Conversely, \papername lifts all instructions to a high-level representation, making it suitable for integration with existing compiler infrastructure and debloating tools. Direct performance comparisons with TOP are not feasible as the source code of TOP is not available. The overhead of TOP is expected to be higher than \papername as TOP relies on QEMU. Finally, the overhead of TOP has been only evaluated against \emph{coreutils} , which are relatively simpler compared to SPEC CPU.

\subsection{Binary Rewriting}

Finally, we discuss the two most recent advances in binary rewriting: Increment CFG patching (ICFGP) \cite{icfg} and Armore~\cite{armore}. ICFGP proposes a number of heuristics and an efficient fallback mechanism for the recovery of indirect branch targets, whereas Armore proposes a low-overhead dynamic resolution of indirect branches in rewritten binaries without using heuristics. At the same time, \papername aims for a heuristic-free indirect control-flow recovery, however, it cannot rely on a dynamic fallback mechanism, as the target of the indirect branch may not be lifted.

\section{Conclusions}

To debloat, specialize or understand a given application binary, it is helpful to be able to lift the executable from binary form into a well-supported and higher-level programming language, such as C, or into an intermediate representation of a compiler. Most available binary debloaters and lifters are focused on x86 architectures, with few examples targeting ARM despite its widespread occurrence.

We have presented a new tool, \papername, the first debloater utilizing a novel hybrid lifting that combines static heuristic-free analysis, instructions lifting and control-flow structuring with dynamic execution tracing, allowing for faster lifting of real-world off-the-shelf applications without relying on unsafe static heuristics or a slow concolic execution. \papername combines the precision of control flow recovery of dynamic lifters, with relatively short lifting times of static lifters. We have shown that the overhead of running binaries under \papername can be as low as 1.05$\times$, with a geomean of $1.78\times$ ($1.40\times$ for the dynamic part). Existing dynamic lifters take hours to lift one complex application and static lifters often fail. We have shown that \papername can debloat complex applications despite them being optimized. At the same time, \papername, unlike the majority of the debloaters, combines the benefits of binary and source code debloating. We showed that the percentage of gadgets of the original binary has a geomean between 24.10\% and 30.22\%, becoming as low as 16.65\% when using ARM PAC pointers. The code size has a geomean as low as 53.59\% when only observed traces are lifted.

\begin{acks}
This research was partially supported by the UK Industrial Strategy Challenge Fund (ISCF) under the Digital Security by Design (DSbD) Programme delivered by UKRI as part of the MoatE (10017512) and Soteria (75243) projects. Mikel Luj\'an is funded by an Arm/RAEng Research Chair award and a Royal Society Wolfson Fellowship.
\end{acks}

\bibliographystyle{ACM-Reference-Format}
\bibliography{draft}


\clearpage
\appendix

\section{Example}

To contextualize the information provided in this paper, we present a simple example illustrating how different inputs generate specific execution traces. Using a basic C program compiled into an assembly language, we can observe how different inputs influence the execution flow and provide different execution traces.

\subsection{C Code}

The following C program serves as the basis for our example:

\begin{lstlisting}[language=C]
int main(int argc, char** argv) {  
    if(argc % 2) {  
        printf("Even!\n");  
    } else {  
        printf("Odd!\n");  
    }  
    return 0;  
}  
\end{lstlisting}

\subsection{AArch64 Assembly}

We first analyze the execution traces generated when the above program is compiled into AArch64 assembly:

\begin{lstlisting}
   main:  

01          stp     x29, x30, [sp, -16]!
02          mov     x29, sp
03          tbz     x0, 0, .L2
04          adrp    x0, .LC0
05          add     x0, x0, :lo12:.LC0
06          bl      puts
    .L3:
07          mov     w0, 0
08          ldp     x29, x30, [sp], 16
09          ret
    .L2:
10          adrp    x0, .LC1
11          add     x0, x0, :lo12:.LC1
12          bl      puts
13          b       .L3
\end{lstlisting}
\medskip

\noindent
\textbf{Example observed trace (argc = 3):}  

\begin{lstlisting}
BB0: (01,03) -> BB1
BB1: (10,12) -> BB2
BB2: (13,13) -> BB3
BB3: (07,09) -> ...
\end{lstlisting}
\medskip

\noindent
\textbf{Example observed trace (argc = 2):}  

\begin{lstlisting}
BB0: (01,03) -> BB4
BB4: (04,06) -> BB3
BB3: (07,09) -> ...
\end{lstlisting}
\medskip

\noindent
\textbf{Merged traces:} 

\begin{lstlisting}
BB0: (01,03) -> BB1, BB4
BB1: (10,12) -> BB2 
BB2: (13,13) -> BB3
BB3: (07,09) -> ...
BB4: (04,06) -> BB3
\end{lstlisting}
\medskip

Each execution trace consists of basic blocks, where each block is defined by its start and end addresses, along with the branch targets. The observed behavior shows different control flow paths based on the input \texttt{argc}, with the program taking different branches for odd and even values of \texttt{argc}.

\subsection{x86 Assembly}

For comparison, we also provide an example of the program compiled into x86 assembly:

\begin{lstlisting}
    main:
01          sub     rsp, 8
02          and     edi, 1
03          je      .L2
04          mov     edi, OFFSET FLAT:.LC0
05          call    puts
    .L3:
06          xor     eax, eax
07          add     rsp, 8
08          ret
    .L2:
09          mov     edi, OFFSET FLAT:.LC1
10          call    puts
11          jmp     .L3  
\end{lstlisting}
\medskip

\noindent
\textbf{Example trace (argc = 3):}  

\begin{lstlisting}
BB0: (01,03) -> BB1
BB1: (09,10) -> BB2
BB2: (11,11) -> BB3
BB3: (06,08) -> ...
\end{lstlisting}
\medskip

In this case, the trace is structurally similar to the AArch64 example. The basic blocks remain consistent, defined by their bounds and branch targets. The main difference lies in the specific assembly instructions that are lifted differently depending on the target architecture, but the control flow remains comparable across both platforms.

\end{document}